# A Block Cipher Using Rotation and Logical XOR Operations

[1]D. Sravan Kumar [2]CH. Suneetha [3]A.Chandrasekhar

[1]Reader in Physics, SVLNS Government College, Visakhapatnam, India

[2]Assistant Professor in Engineering Mathematics, GITAM University, Visakhapatnam, India

[3]Professor in Engineering Mathematics, GITAM University, Visakhapatnam, India

**Abstract**

Cryptography is the study of methods of sending messages in disguised form so that only the intended recipients can remove the disguise and read the messages. Information security has become a very critical aspect of modern communication systems. With the global acceptance of the Internet as a medium of communication, virtually every computer in the world is connected to every other. It has created a new risk for the users of the computers with a constant threat of being hacked and being victims of data theft. In this connection data encryption has become an essential part of secure communication of the messages. In the present paper we propose a new method of encryption of data in blocks using the operations Rotation and Logical XOR.

.

## 1. Introduction

Encryption is the process of converting a plaintext message into cipher text which can be decoded back into the original message using the secret key. An encryption/ decryption algorithm, along with a key is used in the encryption/decryption of data. There are several types of data encryption schemes which form the basis of network security. Encryption schemes are generally based on either block or stream ciphers. Historically the focus of encryption has been on the use of symmetric encryption to provide confidentiality. It is only in the last several decades that other considerations, such as authentication, integrity, digital signature have been included in the theory and practice of cryptology. The security of the message basically depends on two factors 1) confidentiality and 2)Authentication. One of the means of achieving confidentiality of the message is encrypting bulk digital data using block ciphers. Single round of encryption offers inadequate security but multiple rounds offer increasing security. In the present paper we propose a new method of encryption of data block in 8 rounds using Rotation and Logical XOR operations with a one-time sub key derived for each round from the session key of that particular data block which will be generated from the master key(private key). In the key scheduled algorithm [1, 2] for encryption/ decryption of the data proposed in this paper, the size of the data block is selected to be 64 characters. The characters of each data block are coded to 8 bit binary format using ASCII code table and are written as an 8x8 matrix. The binary digits of each element of the message matrix are rotated to new positions so that the outcome is a new element. The number of places an element rotated is different for different elements in each round of encryption i.e. rotation of the digits is not fixed, but depends on the sub keys derived for each round of encryption from the session key of each data block. The session key of each data block is generated [13, 14] from the master key (private key) agreed upon by the communicating parties. Between two successive rotation operations the logical XOR operation is performed on each element of the matrix with its nearest four neighbouring elements so that on completion of eight rounds of rotations and XOR operations good avalanche effect is achieved which is one of the desired properties of encryption algorithm. The procedure designed in this method ensures that the message is highly secure as long as the key selected by the communicating parties is secure. The encryption/ decryption procedure further assures relatively low computation overhead. A Logical XOR gate is digital logical gate which performs a logical operation on one or more logic inputs and produces a single logic output. Several researchers of cryptography used the logical operation XOR [7, 8] in their encryption protocols.

For describing the algorithm the following notation and definitions are adopted:-

## 2. Symbols and Notation

| Symbol | Expression | Meaning |
|---|---|---|
| M,m,n,l | ${}^{l}M_{n}^{m}$ | The 8x8 matrix whose elements are in the 8 bit binary format where l,m,n take integer values, n∈N, l, m ∈ {0,1,2……8] |
| K,m,n | $K_{n}^{m}$ | The 8x8 matrix whose elements are the decimal digits from 0 to 7 |
| R,E, ∧ | $\overset{\wedge}{R}_{E}$ | The rotation operator used in the encryption process. |
| R,D, ∧ | $\overset{\wedge}{R}_{D}$ | The rotation operator used in the decryption process. |
| X,E, ∧ | $\overset{\wedge}{X}_{E}$ | The logical XOR operator used in the encryption process |
| X,D, ∧ | $\overset{\wedge}{X}_{D}$ | The XOR logical operator used in the decryption process |
| S, ∧ | $\overset{\wedge}{S}_{n}^{m}$ | The operator used for deriving the sub key for the $m^{th}$ round of encryption from the key used for the first round encryption of $n^{th}$ data block |
| G, ∧ | $\overset{\wedge}{G}_{n}$ | The operator used for generating the session key for the encryption of the $n^{th}$data block from the session key used in the encryption of the $(n-1)^{th}$data block. |
| M,l,m,n,i,j | $\left[{}^{l}M_{n}^{m}\right]_{ij}$ | Represents the element in the $i^{th}$ row and $j^{th}$ column of the matrix $\left[{}^{l}M_{n}^{m}\right]$ |

## 3. Definitions

1) ${}^{l}M_{n}^{m}$ denotes an 8x8 matrix whose elements are 8 bit binary numbers. The right superscript m denotes the number of rotation operations ( $\overset{\wedge}{R}_{E}$ ) performed on the elements of the message matrix M. The left superscript l represents the number of times the logical XOR operator ( $\overset{\wedge}{X}_{E}$ ) is applied on the message matrix M. The right subscript n indicates the number of data block that is being encrypted.

2) $K_{n}^{m}$ denotes an 8x8 matrix whose elements are the decimal digits ∈{0,1,…7} which is called the key matrix used in the rotation operation of the $(m+1)^{th}$ round encryption of the $n^{th}$ data block matrix ${}^{m}M_{n}^{m}$. The right superscript m of K represents the number of round of encryption of the matrix ${}^{m}M_{n}^{m}$, where m ranges from 0 to 7. The right subscript indicates the data block which is being encrypted.

3). The Operator $\overset{\wedge}{R}_{E}\left[{}^{m}M_{n}^{m},K_{n}^{m}\right]$ is the right rotation operator used in the encryption process. It has two arguments ${}^{m}M_{n}^{m}$ and $K_{n}^{m}$. The first argument is the matrix on which the operation $\overset{\wedge}{R}_{E}$ is applied. The second argument $K_{n}^{m}$ is the key matrix used for performing the operation $\overset{\wedge}{R}_{E}$ in the $(m+1)^{th}$ round encryption of $n^{th}$ data block matrix. With this operation the right superscript m of the matrix ${}^{m}M_{n}^{m}$ increases by one unit. With this operation the digits of each element $\left[{}^{m}M_{n}^{m}\right]_{ij}$ of the matrix ${}^{m}M_{n}^{m}$ which is in the 8 bit binary format are right rotated $\left[K_{n}^{m}\right]_{ij}$ places.

Every element $\left[{}^{m}M_{n}^{m}\right]_{ij}$ consists of 8 bits

$b_7$ $b_6$ $b_5$ $b_4$ $b_3$ $b_2$ $b_1$ $b_0$

$\left[{}^{m}M_{n}^{m}\right]_{ij}$ = $[b_7\ b_6\ b_5\ b_4\ b_3\ b_2\ b_1\ b_0]$ here

$b_i \in \{0,1\}$ and $i \in \{0,1,\ldots.7\}$

The binary digits $b_7$………$b_0$ are right rotated the number of places equal to $\left[K_{n}^{m}\right]_{ij}$

Which results in the matrix ${}^{m}M_{n}^{m+1}$.

$$\overset{\wedge}{R}_{E}\left[{}^{m}M_{n}^{m},K_{n}^{m}\right]={}^{m}M_{n}^{m+1}$$

$\left[K_n^m\right]_{ij} = p$ say

Let the right rotation mapping by p times be denoted by $\overset{\Lambda}{R}_E$ (p). Then the right rotation p times corresponds to the mapping

$$\overset{\Lambda}{R}_E \text{ (p)} : b_j \rightarrow b_{[(j-p) \bmod 8)]}$$

Example let $M_{ij}$ is right rotated three times then

$b_7 \rightarrow b_{(7-3) \bmod 8} = b_4$

$b_6 \rightarrow b_{(6-3) \bmod 8} = b_3$

…………………..

…………………..

$b_1 \rightarrow b_{(1-3) \bmod 8} = b_6$

$b_0 \rightarrow b_{(0-3) \bmod 8} = b_5$

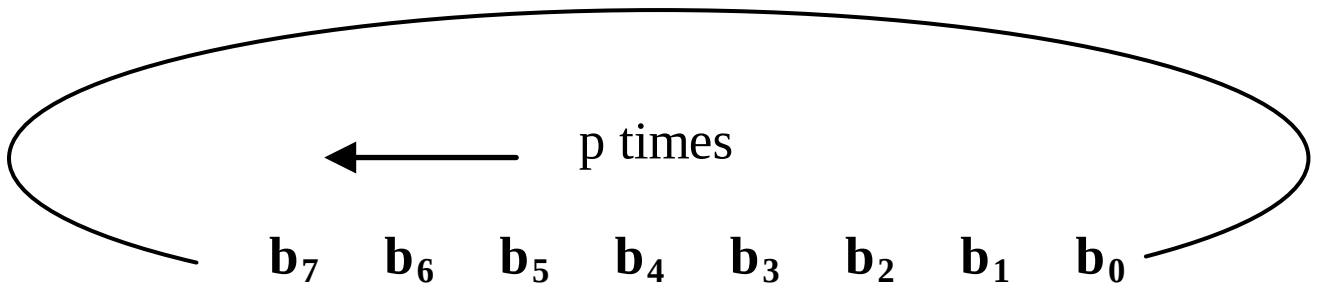

**Example:**

Let $\left[{}^mM_n^m\right]_{24} = 10100100$ and $\left[K_n^m\right]_{24} = 3$ then $\left[{}^mM_n^{m+1}\right]_{24} = 10010100$

4). $\overset{\Lambda}{X}_E[{}^mM_n^{m+1}]$ represents the XORing of each element of the matrix ${}^mM_n^{m+1}$ with the nearest four neighboring elements in the encryption process. With this operation the left superscript m increases by one unit. With this operation each element $\left[{}^mM_n^{m+1}\right]_{ij}$ of the matrix ${}^mM_n^{m+1}$ which is in the 8 bit binary format is XORed with the nearest four neighboring elements which results in the matrix ${}^{m+1}M_n^{m+1}$

$$\overset{\Lambda}{X}_E[{}^mM_n^{m+1}] \rightarrow ((((\,{}^mM_n^{m+1}{}_{ij}\, XOR\, {}^mM_n^{m+1}{}_{ij-1})\; XOR\, {}^mM_n^{m+!}{}_{i+1,j})\, XOR\, {}^mM_n^{m+1}{}_{ij+1})\, XOR\, {}^mM_n^{m+!}{}_{i-1j})$$

$$\overset{\Lambda}{X}_E[{}^mM_n^{m+1}] = {}^{m+1}M_n^{m+1}$$

**Example**

$\left[{}^mM_n^{m+1}\right]_{24} = 10010100$ , $\left[{}^mM_n^{m+1}\right]_{23} = 10100010$

$\left[{}^mM_n^{m+1}\right]_{34} = 00111000$ $\left[{}^mM_n^{m+1}\right]_{25} = 10010010$,

$\left[{}^mM_n^{m+1}\right]_{14} = 01100101$ then

$$\overset{\Lambda}{X}_E\left[{}^mM_n^{m+1}\right]_{24} = ((((10010100\, XOR\, 10100010)\; XOR\, 00111000)\, XOR\, 10010010)\, XOR\, 01100101)$$

$$\left[{}^{m+1}M_n^{m+1}\right]_{24} = 11111001$$

5). $\overset{\Lambda}{X}_D[{}^mM_n^m]$ represents the XORing of each element of the matrix ${}^mM_n^m$ with the surrounding elements in the decryption process. With this operation the left superscript m decreases by one unit.

$$\overset{\Lambda}{X}_D[{}^mM_n^m] \rightarrow ((((\,{}^mM_n^m{}_{ij}\, XOR\, {}^mM_n^m{}_{i-1j})\; XOR\, {}^mM_n^m{}_{ij+1})\, XOR\, {}^mM_n^m{}_{i+1j})\, XOR\, {}^mM_n^m{}_{ij-1})$$

. $\overset{\Lambda}{X}_D[{}^mM_n^m] = {}^{m-1}M_n^m$ where m ranges from 0 to 7.

**Example**

$\left[{}^mM_n^m\right]_{45} = 11111001$ $\left[{}^mM_n^m\right]_{35} = 10100010$

$\left[{}^mM_n^m\right]_{46} = 00111000$ $\left[{}^mM_n^m\right]_{55} = 10010010$

$\left[{}^mM_n^m\right]_{44} = 01100101$ Then

$$\overset{\Lambda}{X}_D\left[{}^mM_n^m\right]_{45} = ((((11111001\, XOR\, 01100101)\; XOR\, 10010010)\, XOR\, 00111000)\, XOR\, 10100010)$$

$$\left[{}^{m-1}M_n^m\right]_{45} = 10010100$$

6). The Operator $\overset{\wedge}{R}_D\left[{}^{m-1}M_n^m, K_n^m\right]$ is the left rotation operator used in the decryption process. It has two arguments ${}^{m-1}M_n^m$ and $K_n^m$. The first argument is the matrix on which the operation $\overset{\wedge}{R}_D$ is performed in the (9-m)$^{th}$ round decryption of the nth data block matrix. The second argument $K_n^m$ is the key matrix used for performing the operation $\overset{\wedge}{R}_D$ in the m$^{th}$ round decryption of the n$^{th}$ data block matrix. With this operation the right superscript m of the matrix ${}^{m-1}M_n^m$ decreases by one unit. With this operation the digits of each element $\left[{}^{m-1}M_n^m\right]_{ij}$ of the matrix ${}^{m-1}M_n^m$ which is in the 8 bit binary format are right rotated $\left[K_n^m\right]_{ij}$ places.

i,e, every element $\left[{}^{m-1}M_n^m\right]_{ij}$ consists of 8 bits

$b_7$ $b_6$ $b_5$ $b_4$ $b_3$ $b_2$ $b_1$ $b_0$

$$\left[{}^{m-1}M_n^m\right]_{ij} = [b_7\, b_6\, b_5\, b_4\, b_3\, b_2\, b_1\, b_0]$$

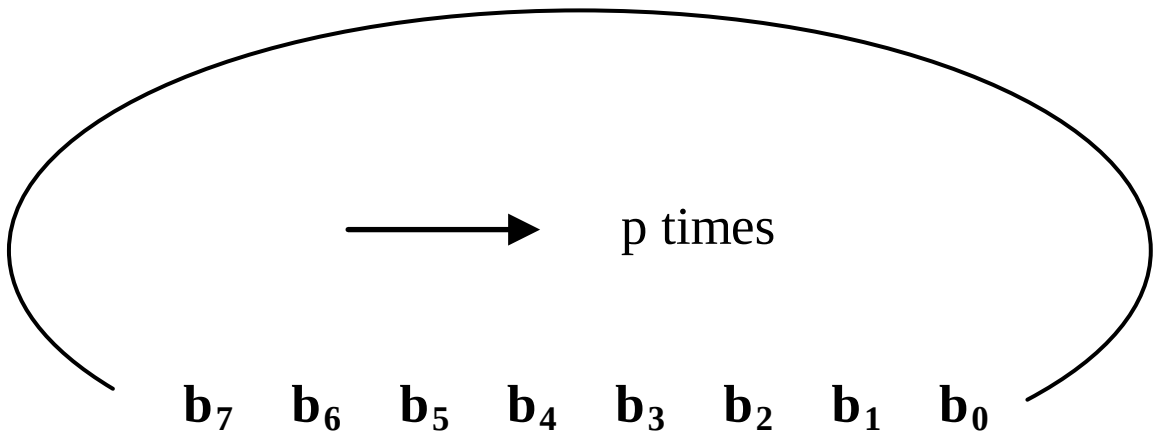

the binary digits $b_7$………$b_0$ are left rotated the number of times equal to $\left[K_n^m\right]_{ij}$

which results in the matrix ${}^{m-1}M_n^{m-1}$.

$$\overset{\wedge}{R}_D\left[{}^{m-1}M_n^m, K_n^m\right] = {}^{m-1}M_n^{m-1}$$

7). $\overset{\wedge}{S}{}_n^m$ is the operator used for deriving the sub key for the (m+1)$^{th}$ round encryption of nth data block from the session key used for the first round operation i.e.

$$\overset{\wedge}{S}{}_n^m[K_n^0] = K_n^{m-1}$$

The key matrix $K_n^{m-1}$ for the m$^{th}$ round encryption of the nth data block is obtained from $K_n^0$ by shifting the columns of the matrix $K_n^0$ to the right by (m-1) places.

i.e. $\left[K_n^{m-1}\right]_{ij} = \left[K_n^0\right]_{ij-m+1}$

8). $\overset{\wedge}{G}_n$ is the operator which defines the session key generation for the first round encryption of the nth data block from the session key used for the first round encryption of the (n-1)$^{th}$ data block.

$$\overset{\wedge}{G}_n\left[K_{n-1}^0\right] = K_n^0, where\left[K_n^0\right]_{ij} = \left[(K_{n-1}^0)_{ij} + (K_{n-1}^0)_{ij+1}\right]_{\bmod 8}$$

i **and j take values in all the definitions made above from 0 to 7. If i+1 or i-1 or j+1 or j-1 or any subscript fall out of the range {0,1,2,….7} then modulo 8 of that number be considered.** The session key of each data block is itself sub key for the first round of encryption of the data block.

Before communicating the messages both the sender and the receiver agree upon to use the secret key which is in the form of an 8x8 matrix K (master key) whose elements are the decimal digits from 0 to 7. This matrix K (master key) is denoted by $K_1^0$ in the encryption/ decryption process i.e. the master key itself is the session key for the encryption/decryption of first data block. For implementing the algorithm the entire message is divided into data blocks of 64 characters each $D_1, D_2, D_3$……$D_n$ where n is a natural number. The characters in each message block are coded to 8 bit binary numbers using ASCII code table and are arranged in the form of 8x8 matrices ${}^0M_1^0, {}^0M_2^0, {}^0M_3^0, \ldots\ldots {}^0M_n^0$ row wise. The number of characters in the message always may not be the integral multiple of 64. Hence, at the end of the message the sender adds three # characters (###) and ensures that the message fills integer number of text blocks by adding random different characters after the three # characters.

## 4. Algorithm

### 4.1 Encryption-

K = Key agreed upon by the communicating parties

Set n = 1

m = 1

Step 1:- PRINT DATA BLOCK = n

If n = 1 $K_n^0 = K$

else

$$\hat{G}_n\left[K_{n-1}^0\right] = K_n^0,\ where\left[K_n^0\right]_{ij} = \left[(K_{n-1}^0)_{ij} + (K_{n-1}^0)_{ij+1}\right]_{\mathrm{mod}8}$$

Step2:- PRINT ENCRYPTION ROUND = m

Step3:- $S_n^m[K_n^0] = K_n^{m-1}$

where $\left[K_n^{m-1}\right]_{ij} = \left[K_n^0\right]_{ij-m+1}$

Step4:- $\hat{R}_E\left[{}^{m-1}M_n^{m-1}, K_n^{m-1}\right] = {}^{m-1}M_n^m$

Step5:- $\hat{X}_E[{}^{m-1}M_n^m] = {}^mM_n^m$

Step6:- If m < 8, increment m by one unit and go to Step 2

Else set m = 1

If n < N increment n by one unit and go to Step 1.

Else Stop

After 8 rounds of encryption using rotation operation and logical XOR operation the encrypted message matrices ${}^8M_1^8, {}^8M_2^8, {}^8M_3^8 ....... {}^8M_n^8$ are obtained. Then all the elements of the matrices which are in 8 bit binary format are converted into equivalent text characters using ASCII code table to get the cipher data blocks $D_1^E, D_2^E, D_3^E ......., D_n^E$. This cipher text is communicated to the receiver through the public channel.

## 4.2 Decryption

The receiver after receiving the cipher text divides the message into data blocks $D_1^E, D_2^E, D_3^E ......., D_n^E$ of 64 characters each. All the 64 characters of each data block are converted into 8 bit binary numbers using ASCII code table and all the elements are written as 8x8 matrices ${}^8M_1^8, {}^8M_2^8, {}^8M_3^8 ....... {}^8M_n^8$

K = Key agreed upon by the communicating parties

Set n = 1

m = 8

Step 1:- PRINT DATA BLOCK = n

If n = 1 $K_n^0 = K$

Else

$$\hat{G}_n\left[K_{n-1}^0\right] = K_n^0,\ where\left[K_n^0\right]_{ij} = \left[(K_{n-1}^0)_{ij} + (K_{n-1}^0)_{ij+1}\right]_{\mathrm{mod}8}$$

Step2:- PRINT DECRYPTION ROUND = 9-m

Step3:- $S_n^m[K_n^0] = K_n^{m-1}$,

Where $\left[K_n^{m-1}\right]_{ij} = \left[K_n^0\right]_{ij-m+1}$

Step4:- $\hat{X}_D[{}^mM_n^m] = {}^{m-1}M_n^m$

Step5:- $\hat{R}_D\left[{}^{m-1}M_n^m, K_n^{m-1}\right] = {}^{m-1}M_n^{m-1}$

Step6:- If m > 0, decrement m by one unit and go to Step 2

Else set m = 8

If n < N increment n by one unit and go to Step 1

Else Stop

After 8 rounds of decryption using logical XOR operation and rotation operation the original message matrices ${}^0M_1^0, {}^0M_2^0, {}^0M_3^0, ...... {}^0M_n^0$ are obtained. Then all the elements of the matrices ${}^0M_1^0, {}^0M_2^0, {}^0M_3^0, ...... {}^0M_n^0$ which are in 8 bit binary format are converted into text characters using ASCII code table to get the original message blocks $D_1, D_2, D_3 \ldots\ldots D_n$

## 5. Security Analysis

In the key scheduled algorithm proposed here different keys are used for encrypting different data blocks which are called session keys generated from the master key (secret key between the sender and the receiver) and the key used for the encryption of each round is different and is derived from the session key of the corresponding round which is called the sub key. As different keys are used for different data blocks cipher is less vulnerable to passive attacks. As each element of the message matrix M is rotated (not fixed rotation) and logical XOR operation is performed with its all nearest neighboring elements the same characters in the plain text space are mapped to different characters of the cipher text space even though they are in the same text block or different text blocks. So, cipher text is not easily amenable to cryptanalysis [6, 7]. Even the change of a single element of the message matrix changes almost the entire cipher block matrix, i.e., to say that the proposed algorithm has achieved a good avalanche effect [4, 5] which is one of the desired qualities of a good encryption algorithm.

If the same message is sent in I and II (or any subsequent) data block, they are mapped to different cipher texts, i.e., even if the same message is sent repeatedly in the same message block, the messages are enciphered to different cipher texts. Hence, active attacks such as chosen plain text attacks [6, 15], chosen cipher text attacks [9, 10, 16] are quite difficult to execute. Hence, the proposed algorithm is less vulnerable to active attacks. The present encryption algorithm is at most secure against man-in-middle attack [3, 11, 12] because the entire master key is agreed upon by the sender and the receiver rather than the electronic exchange of the parts of the key.

The proposed key scheduled algorithm in this paper is less prone to timing attacks because the time required to encipher or decipher a data block is same for all data blocks since time for enciphering or deciphering is independent of characters in the data block. Even though the original message contains less that 64 characters the remaining characters are filled at random, so that each data block contains exactly 64 characters.

The size of the key is 64 decimal digits where each decimal digit takes values from 0 to 7. Hence $64^8$ different keys are possible. It is estimated that on a 4GHz single core processor the time required to encipher/ decipher a text block is 18µsec. Hence, the vulnerability to brute force attack is very less [the life time of a human being i.e., 100years is approximately equal to 3Gsec, the time required to try all possible keys to decipher a single cipher block by brute force method is roughly 5Gsec]

It is estimated that the time required to encipher a text book containing 500 pages, each page having 40 lines and each lines having 40 characters is 7½minutes.

Dr. D. Sravana Kumar is a senior faculty in Physics in Government College, Visakhapatnam. He obtained his doctorate in Ultrasonics. His research interest includes Ultrasonics, Molecular Interactions and Cryptography. He has published 12 research papers in various international journals those published by Springer, Elsevier, etc. He is an ex-scientist in the Department of Atomic Energy, Government of India. He is very much interested in interdisciplinary research.

CH. Suneetha is Assistant Professor in Engineering Mathematics in GITAM University, Visakhapatnam. She obtained her master's degree in applied mathematics, M.Phil. in algebra. At present she is pursuing her Ph.D under the guidance of Dr. A. Chandrasekhar. Her research interests include Cryptography, Linear Transformations and Algebraic Curves.

Dr. A. Chandrasekhar is Professor and Head of the Department of Engineering Mathematics in GITAM University, Visakhapatnam. He obtained his doctorate in Cryptography. His research interest includes Cryptography and Fixed Point Theory. He has published 14 research papers in various reputed national and international journals including IEEE.